\documentclass[submission, Proceedings]{SciPost}

\usepackage{siunitx}
\usepackage{subfig}

\newcommand{\mueee}{\mbox{$\mu\,\rightarrow\,eee$}}
\newcommand{\muposeee}{$\mu^+\,\rightarrow\,e^+e^-e^+$}
\newcommand{\mueeemath}{\mu\,\rightarrow\,eee}
\newcommand{\muposeeemath}{\mu^+\,\rightarrow\,e^+e^-e^+}
\newcommand{\mueeenunu}{$\mu\,\rightarrow\,eee\nu\nu$}
\newcommand{\muposeeenunu}{$\mu^+\,\rightarrow\,e^+e^-e^+\overline{\nu}_\mu\nu_e$}
\newcommand{\muposeeenunumath}{\mu^+\,\rightarrow\,e^+e^-e^+\overline{\nu}_\mu\nu_e}

\newcommand{\mueX}{$\mu\,\rightarrow\,eX$}

\newcommand{\familon}{$X$}

\newcommand{\DPmath}{A^\prime}
\newcommand{\ee}{$e^+e^-$}
\newcommand{\eemath}{e^+e^-}

\newcommand{\BR}{\text{BR}}

\newcommand{\SM}{Standard Model}
\newcommand{\geant}{GEANT4}
\newcommand{\phase}{\mbox{phase I}}

\DeclareSIUnit[]\muon{\ensuremath{\mu}}

\begin{document}

\begin{center}{\Large \textbf{
The Rare and Forbidden:
Testing Physics Beyond the Standard Model with Mu3e
}}\end{center}

% TODO: write the author list here. Use initials + surname format.
% Separate subsequent authors by a comma, omit comma at the end of the list.
% Mark the corresponding author with a superscript *.
\begin{center}
A. K. Perrevoort\textsuperscript{1*},
on behalf of the Mu3e Collaboration\textsuperscript{2}
\end{center}

\begin{center}
{\bf 1} NIKHEF, Amsterdam, Netherlands (formerly Physics Institute,
Heidelberg, Germany)
\\
{\bf 2} Paul Scherrer Institute, Villigen, Switzerland
\\
* ann-kathrin.perrevoort@cern.ch
\end{center}

\begin{center}
\today
\end{center}

\definecolor{palegray}{gray}{0.95}
\begin{center}
\colorbox{palegray}{
  \begin{tabular}{rr}
  \begin{minipage}{0.05\textwidth}
    \includegraphics[width=8mm]{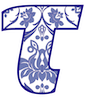}
  \end{minipage}
  &
  \begin{minipage}{0.82\textwidth}
    \begin{center}
    {\it Proceedings for the 15th International Workshop on Tau Lepton Physics,}\\
    {\it Amsterdam, The Netherlands, 24-28 September 2018} \\
    \href{https://scipost.org/SciPostPhysProc.1}{\small \sf scipost.org/SciPostPhysProc.Tau2018}\\
    \end{center}
  \end{minipage}
\end{tabular}
}
\end{center}

% For convenience during refereeing: line numbers
%\linenumbers

\section*{Abstract}
{\bf
The upcoming Mu3e experiment aims to search for the lepton flavour violating
decay $\boldsymbol{\muposeeemath}$ with an unprecedented final sensitivity of
one signal decay in $\boldsymbol{\num{e16}}$ observed muon decays by making
use of an innovative experimental design based on novel ultra-thin silicon
pixel sensors. 
In a first phase, the experiment is operated at an existing muon beam line
with rates of up to $\boldsymbol{\num{e8}}$ muons per second. Detailed
simulation studies confirm the feasibility of background-free operation and
project single event sensitivities in the order of $\boldsymbol{\num{e-15}}$
for signal decays modelled in an effective field theory approach. \\
The precise tracking of the decay electrons and large geometric and momentum
acceptance of Mu3e enable searches for physics beyond the Standard Model in
further signatures. Examples of which are searches for lepton flavour
violating two-body decays of the muon into an electron and an undetected boson
as well as for electron-positron resonances in
$\boldsymbol{\muposeeenunumath}$ which could result for instance from a dark
photon decay. The Mu3e experiment is expected to be competitive in all of
these channels already in phase I. 
}

\vspace{10pt}
\noindent\rule{\textwidth}{1pt}
\tableofcontents\thispagestyle{fancy}
\noindent\rule{\textwidth}{1pt}
%\vspace{10pt}

%% \section{Introduction}
%% \label{sec:intro}
\begin{figure}[h]
\centering
\subfloat[Neutrino mixing. ]{\includegraphics[width=0.3\textwidth, clip,trim=0.7cm 0.3cm 0.7cm 0.25cm]{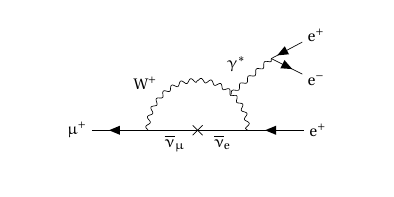}\label{fig:numixing}}\quad
\subfloat[Supersymmetry. ]{\includegraphics[width=0.3\textwidth, clip, trim= 0.7cm 0.35cm 0.7cm 0.25cm]{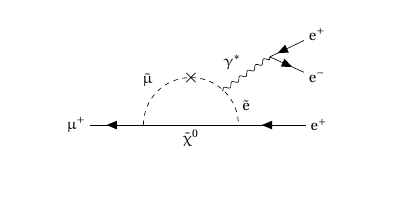}\label{fig:susy}}\quad
\subfloat[Extended electroweak sector. ]{\includegraphics[width=0.3\textwidth, clip, trim= 0.4cm 1cm 0.4cm 0.25cm]{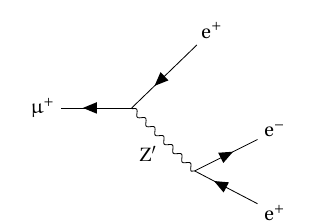}\label{fig:Zprime}}
\caption{The decay \mueee\ via neutrino mixing and in other extensions of the \SM.}
\label{fig:MuLFV}
\end{figure}

\section{Introduction}
\label{sec:intro}
The flavour of leptons is conserved in the Standard Model 
but -- as demonstrated by the observation of neutrino oscillations -- 
it is not conserved in nature. 
The violation of the flavour of charged leptons has however eluded
observation so far. \\
One example for charged lepton flavour violation is the decay \muposeee. 
In a \SM\ extended to include neutrino mixing, it can be mediated in loop
diagrams (see figure~\ref{fig:numixing}) but it is suppressed to branching
fractions below \num{e-54}\ and thus far beyond what experiments can observe. 
Any observation of \mueee\ would therefore be a clear sign for
physics beyond the \SM. 
Indeed, many extensions of the \SM\ predict enhanced rates for \mueee, for
example via loop diagrams with supersymmetric particles (see
figure~\ref{fig:susy}) or at tree-level via a $Z^\prime$ (see
figure~\ref{fig:Zprime}).

\section{The Mu3e Experiment}
\label{sec:mu3e}
The latest measurement of \muposeee\ has been performed by the SINDRUM
experiment~\cite{Bellgardt:1987du}. As no signal was observed, branching
fractions of larger \num{1.0e-12}\ were excluded at \SI{90}{\percent}\
confidence limit (CL). 
The upcoming Mu3e experiment at the Paul Scherrer Institute (PSI) 
will repeat this search with a sensitivity of about one signal
decay in \num{e15}\ muon decays in the first phase of the experiment, 
and ultimately one signal decay in \num{e16}\ muon decays in the
second phase,
thus improving the existing limit by four orders of
magnitude~\cite{Blondel:2013ia}.

\subsection{Signal and Background}
\label{sec:signalBG}
The signal decay \muposeee\ is characterised by the coincidence of two
positrons and one electron\footnote{In the following, the two positrons and
the electron will simply be referred to as three electrons. }\ 
from a common vertex. 
As muons are stopped in the detector and decay at rest, the momenta of
the three decay products vanishes whereas the invariant mass equals the muon
rest mass. \\
There are two sources of background to \mueee\ searches. 
On the one hand, there is background from the rare muon decay \muposeeenunu\
which has the same visible final state as \muposeee. This source of background
can be distinguished from signal by measuring the missing energy carried by
the undetected neutrinos. \\
On the other hand, high muon stopping rates give rise to accidental
combinations of two positrons and an electron from different processes. The
most common source of this background stems from positrons that undergo Bhabha
scattering in the detector material and have a considerable momentum transfer
to the electron. 
Paired with another positron from a close-by Michel decay, these three
particles can mimic a signal decay. 
In addition to kinematic constraints, this type of background is suppressed by
means of vertexing and timing. \\
Thus, for the Mu3e experiment a very accurate electron tracking and a precise
timing measurement are required in addition to high muon stopping rates.

\subsection{Experimental Concept}
\label{sec:detector}
\begin{figure}
\begin{centering}
        \includegraphics[width=0.3\textwidth]{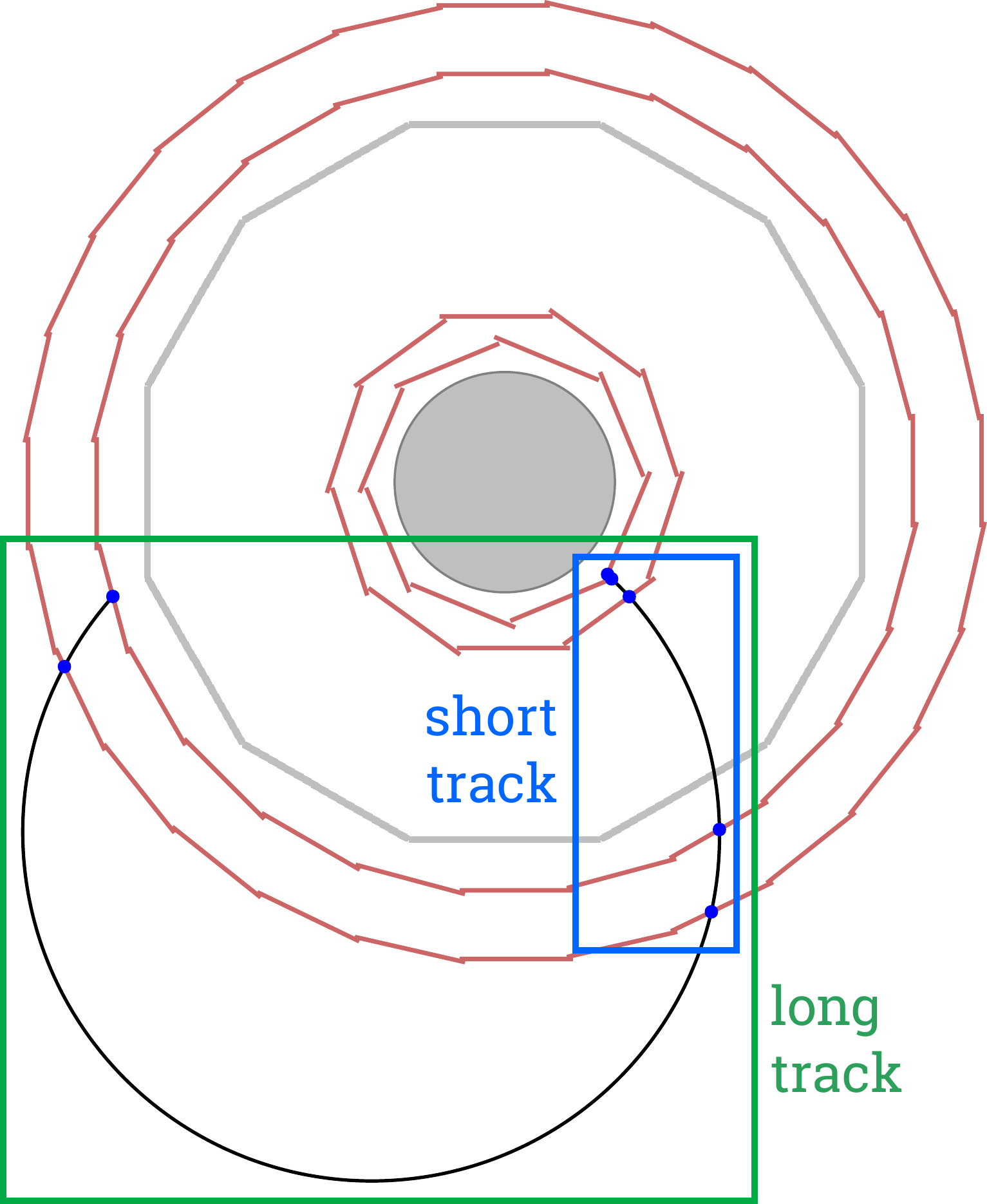} \quad 
        \includegraphics[width=0.6\textwidth]{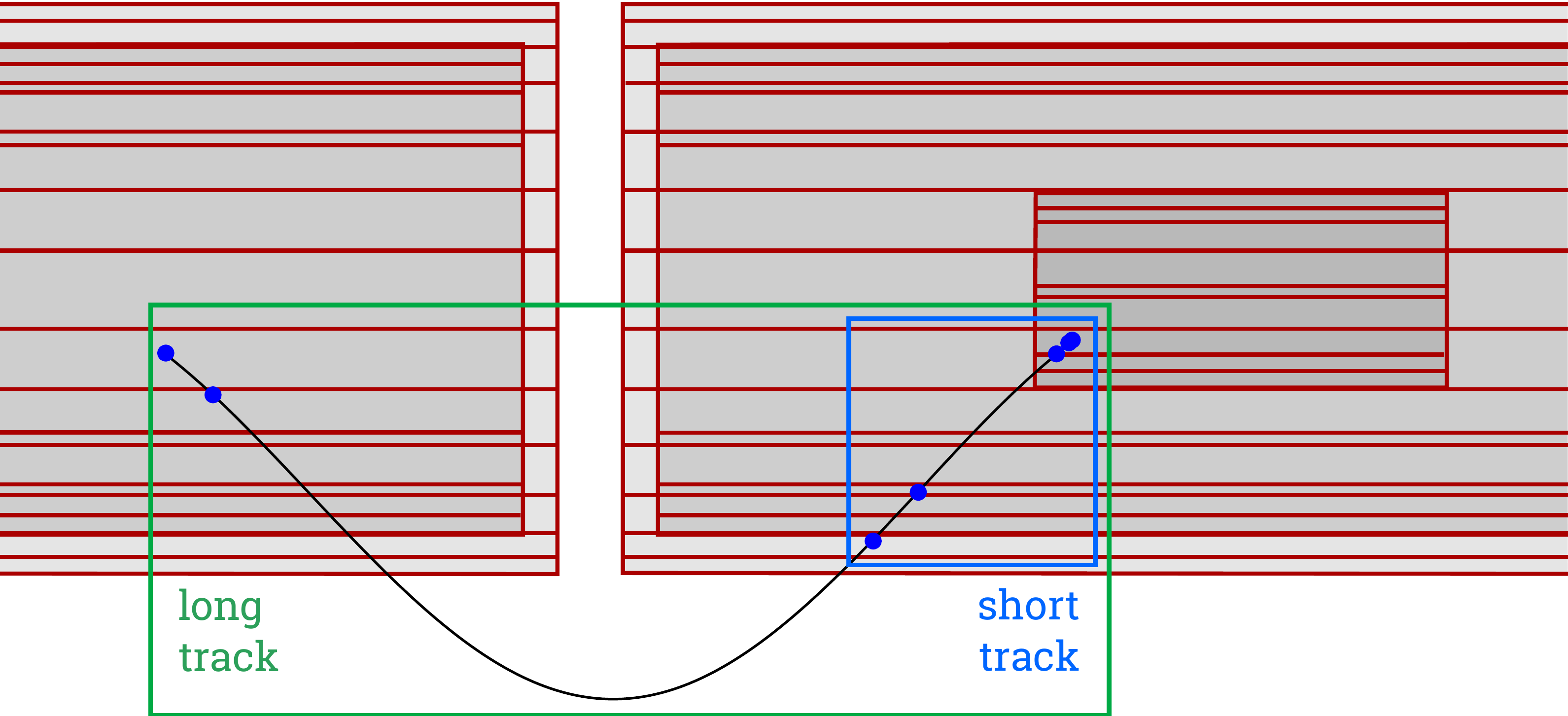} 
\caption{Event display of a simulated recurling track shown in
        transverse and longitudinal projection. The definitions of a
        reconstructed short and long track are shown.}
\label{fig:Mu3eRecurl}
\end{centering}
\end{figure}

\begin{figure}
\begin{centering}
        \subfloat[Short tracks.]{\includegraphics[width=0.45\textwidth]{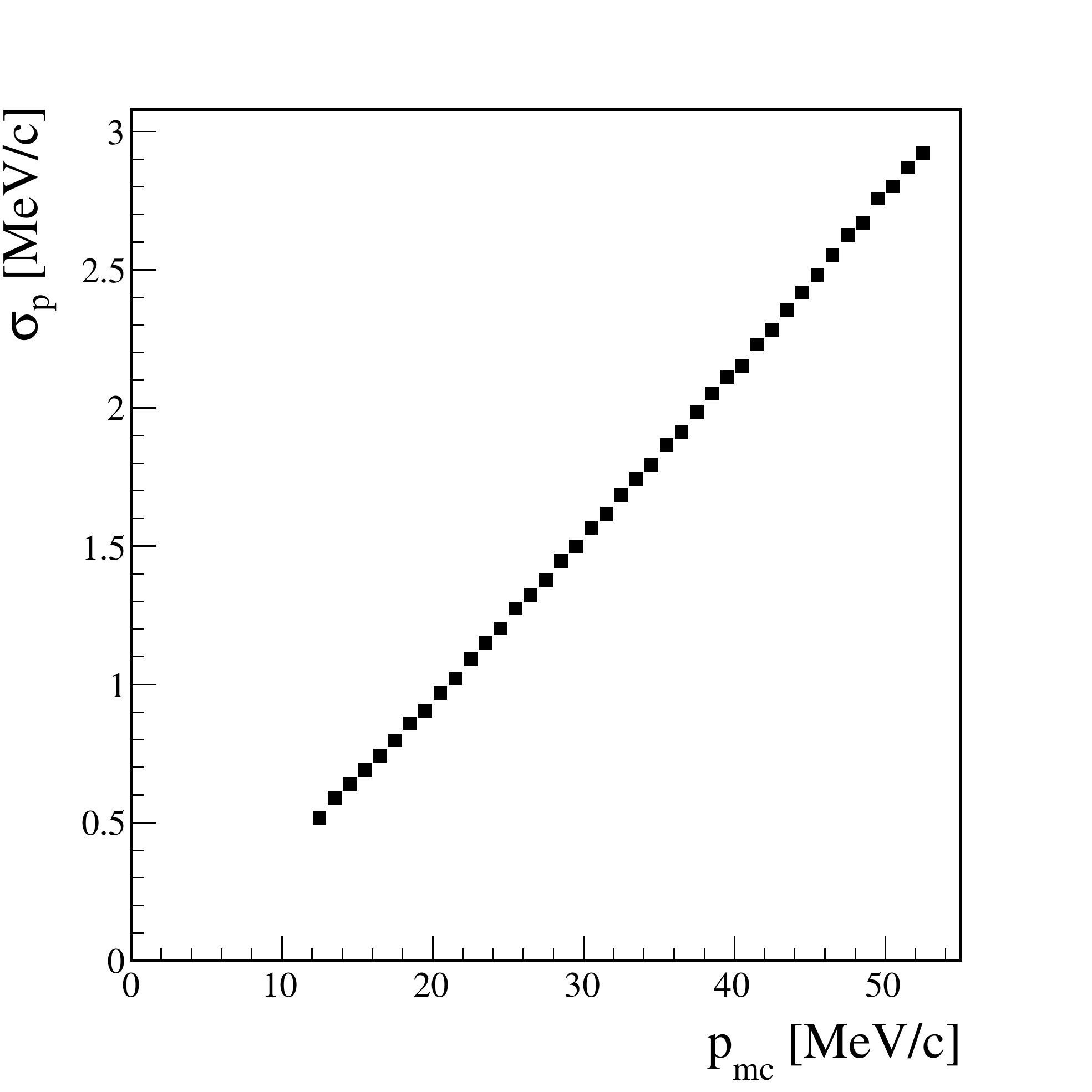}} \quad
        \subfloat[Long tracks. ]{\includegraphics[width=0.45\textwidth]{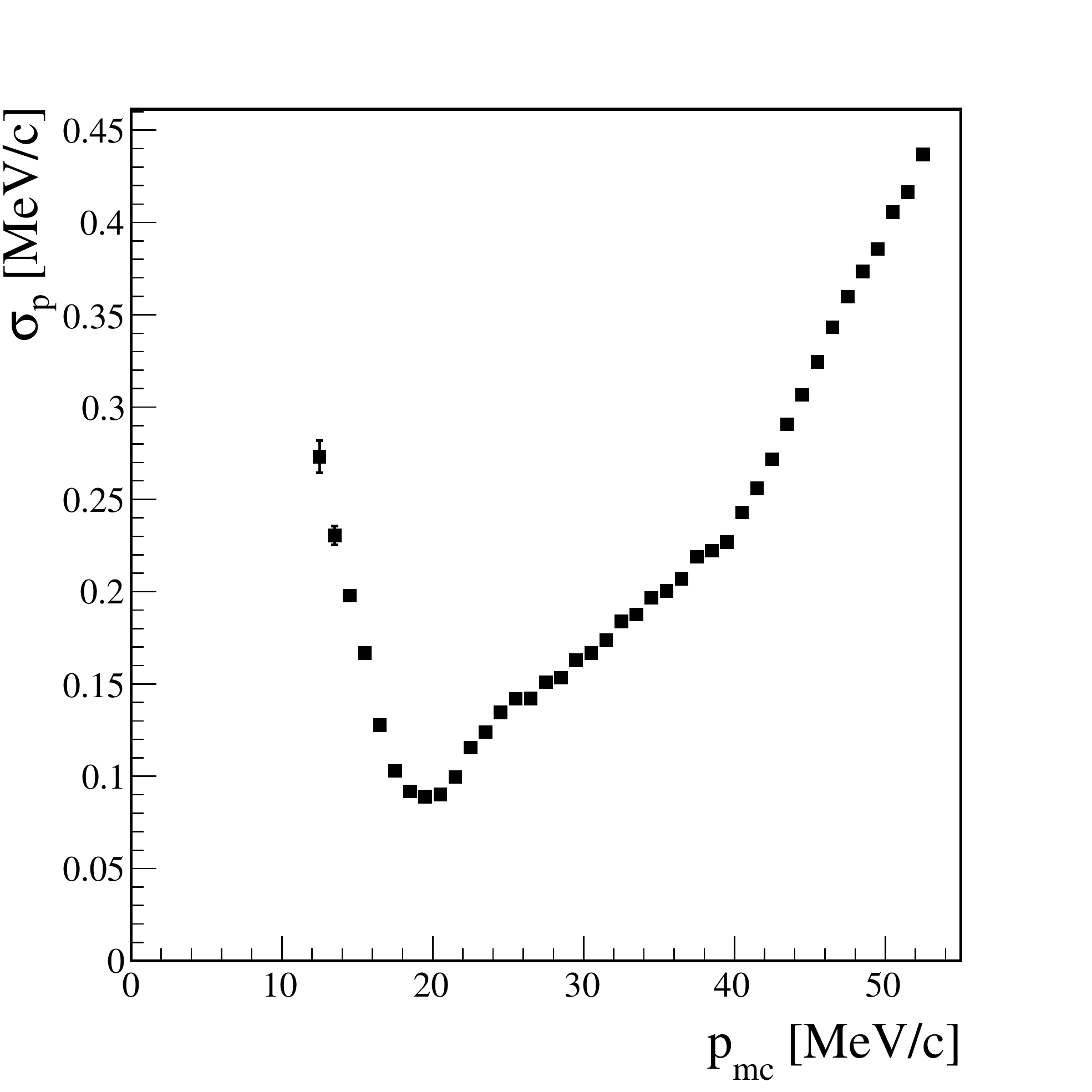}}
\caption{Momentum resolution of reconstructed short and long tracks. }
\label{fig:Mu3eReso}
\end{centering}
\end{figure}

The Mu3e experiment is designed to measure low momentum electrons
with outmost precision and at high rates. 
In \phase, it will be located at an existing muon beam line at PSI which can
provide muon rates of about \SI{e8}{\muon\per\second}. 
In the second phase, higher muon rates in the range
of \SI{e9}{\muon\per\second}\ are required. PSI is currently investigating on
options for the High Intensity Muon Beam line with muon rates of the order
of \SI{e10}{\muon\per\second}. \\
\begin{figure}
\begin{centering}
        \includegraphics[width=\textwidth, clip, trim= 0cm 4.5cm 0cm 3cm]{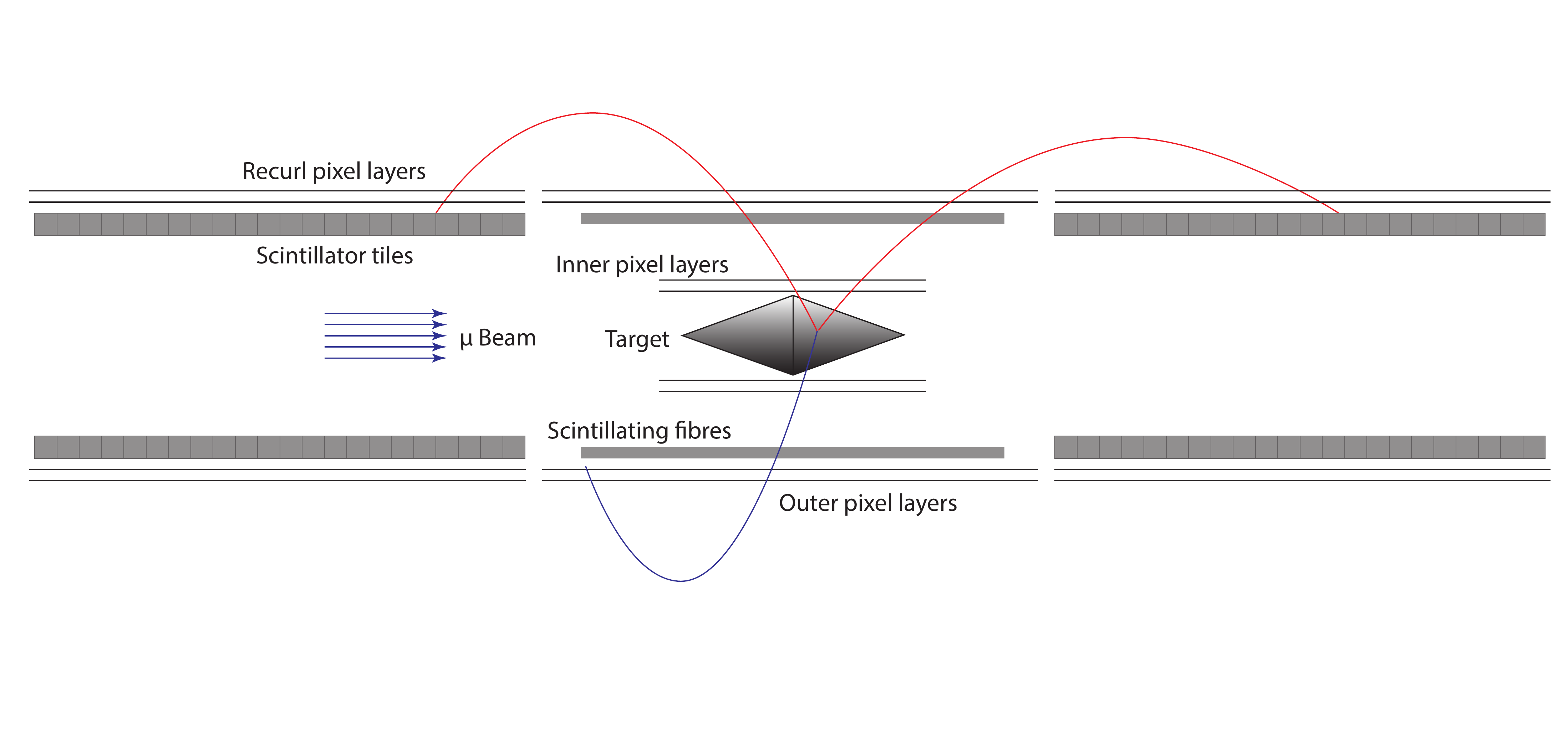}
\caption{Schematic of the phase I Mu3e experiment in lateral projection with a
simulated \mueee\ event. }
\label{fig:Mu3eDet}
\end{centering}
\end{figure}
The detector is shown in figure~\ref{fig:Mu3eDet}. 
The incoming muon beam is stopped on a hollow double cone target made from
Mylar foil. 
The momentum of the decay products is measured by their curvature in the
solenoidal magnetic field. 
The tracking detector is based on silicon pixel sensors and is cylindrically
arranged around the beam axis.  
Two layers of pixel sensors surround the target for precise vertexing while
two further layers are placed at a larger radius for momentum measurements. 
A minimum transverse momentum $p_\text{T}$ of \SI{10}{\MeV}\ is required for a
particle to cross all four layers and be reconstructed. 
A scintillating fibre detector provides precise timing. \\
The electrons from muon decays have low momenta of only a few
tens of \si{MeV}. Hence, multiple Coulomb scattering dominates the momentum
resolution. 
For this reason, the HV-MAPS (High Voltage Monolithic Pixel Sensor) technology
is chosen for the pixel detector~\cite{Peric:2007zz}. 
This technology features a built-in readout circuitry eliminating the need for
additional readout chips.  
HV-MAPS further allow for a sensor thickness of only \SI{50}{\micro\metre}\
and thus a material amount of \SI{0.1}{\percent}\ of a radiation length per
pixel layer including mechanical support and readout flexprint. \\
The precision of the momentum measurement can be further improved by measuring
the trajectory of the electron not only when it is outgoing but also when it
returns to the detector (see figure~\ref{fig:Mu3eRecurl}). 
The momentum resolution of a reconstructed long track from a \emph{recurling}\
particle is up to factor of ten smaller than in the case of reconstructed
short tracks from the outgoing trajectory alone (see
figure~\ref{fig:Mu3eReso}). 
The Mu3e detector is therefore extended by one (two) so-called recurl stations
upstream and downstream of the central detector in \phase\,(phase II). 
The recurl stations consist of two layers of pixel sensors and scintillating
tiles. \\
The data acquisition is performed without a hardware trigger. 
All sub-detectors are continuously read out. 
The data rate is however too high for mass storage so that 
\mueee\ candidates already need to be identified online. 
This is performed on the online filter farm which is based on graphics
processing units. First, short tracks in the central detector are
reconstructed in a fast triplet fit~\cite{Berger:2016vak}. In the following
vertex fit, \mueee\ candidates are looked for and if found all data of the
corresponding readout frame is written to mass storage. 
Reconstruction of long tracks and refined vertex fits are then performed
offline.

\section{Sensitivity Studies}
\label{sec:sensitivity}
The following sensitivity studies are performed with a detailed \geant-based
simulation of the Mu3e detector in \phase. 
The muon stopping rate is assumed to be \SI{e8}{\muon\per\second}, and the
total run time is estimated with 300 data taking days which accumulates to a
total of \num{2.6e15}\ muon decays.

\subsection[$\mueeemath$ in Effective Theories]{$\boldsymbol{\mueeemath}$ in Effective Theories}
\label{sec:EFT}
\begin{figure}[t]
\begin{centering}
        \includegraphics[width=0.75\textwidth]{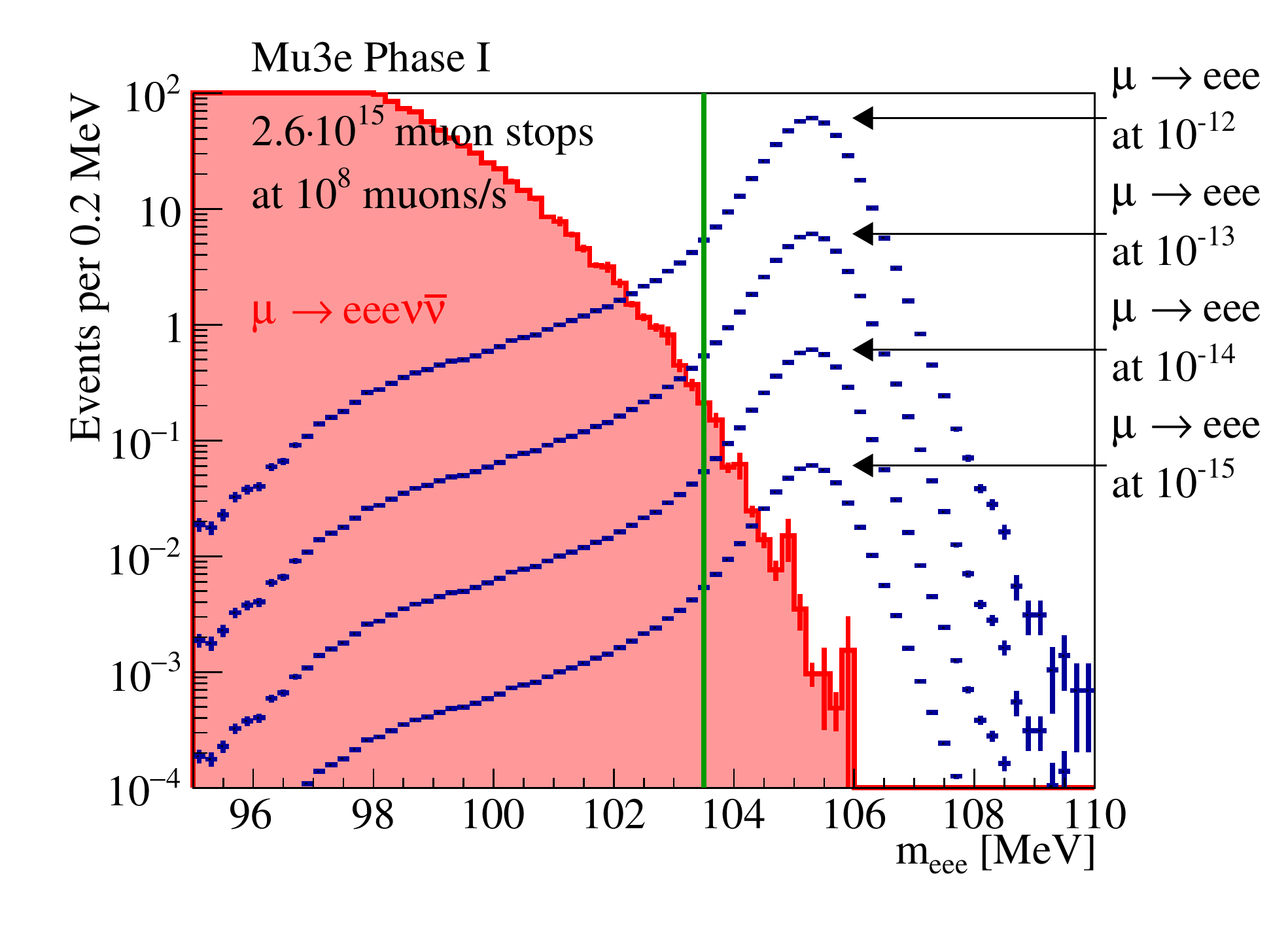}
\caption{Expected distribution of the reconstructed invariant mass of \mueee\
candidates. Background from \mueeenunu\ decays is shown in red. The signal
distribution (blue) is given for various assumptions on the \mueee\ branching
fraction. The applied cut on the three electron mass is indicated by the green
line. }
\label{fig:Mu3e}
\end{centering}
\end{figure}

\begin{figure}[t]
\begin{centering}
        \subfloat[Generated three-body \mueee\ decays. ]{\includegraphics[width=0.3\textwidth]{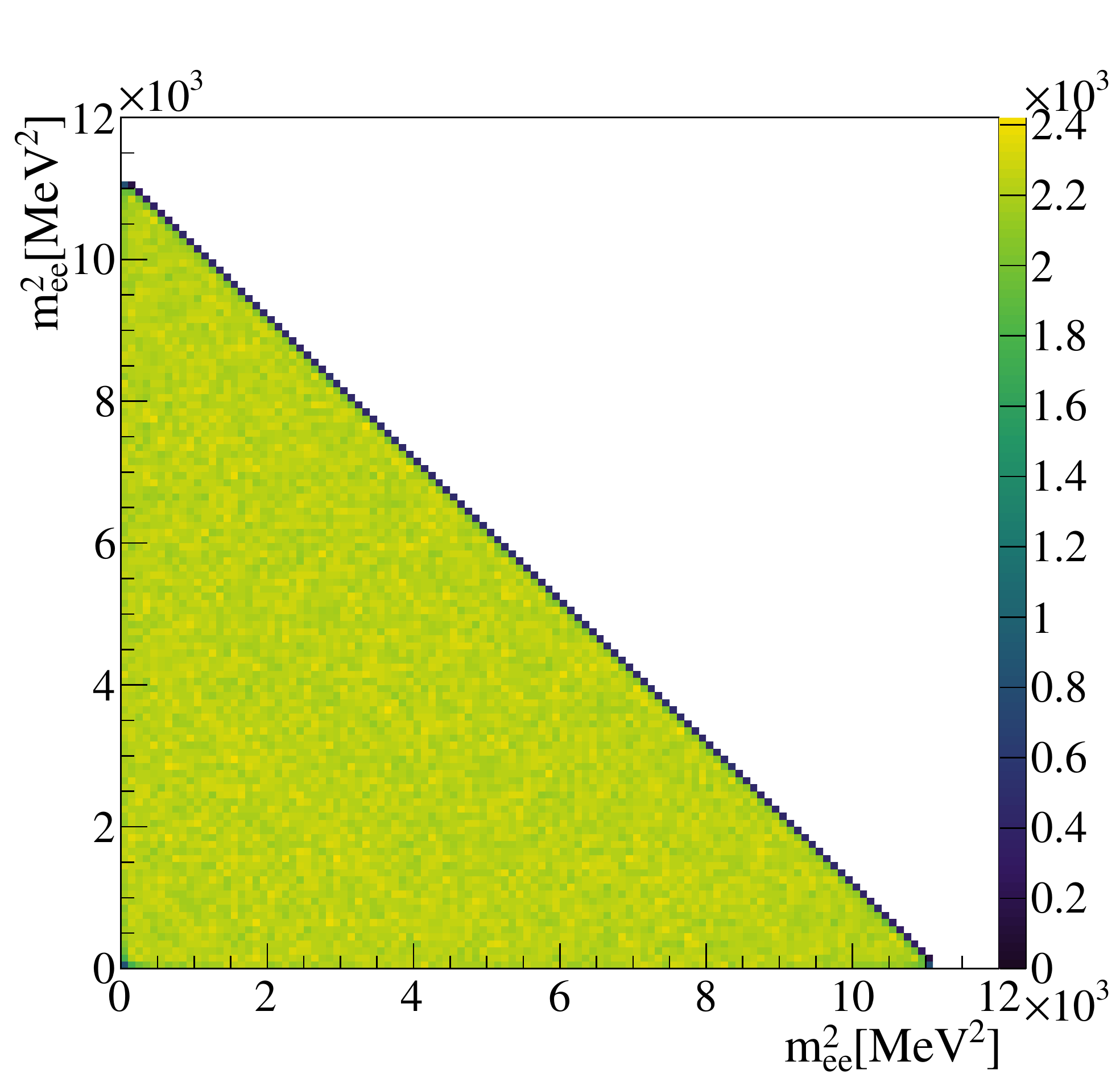}} \quad
        \subfloat[Reconstructed three-body \mueee\ decays. ]{\includegraphics[width=0.3\textwidth]{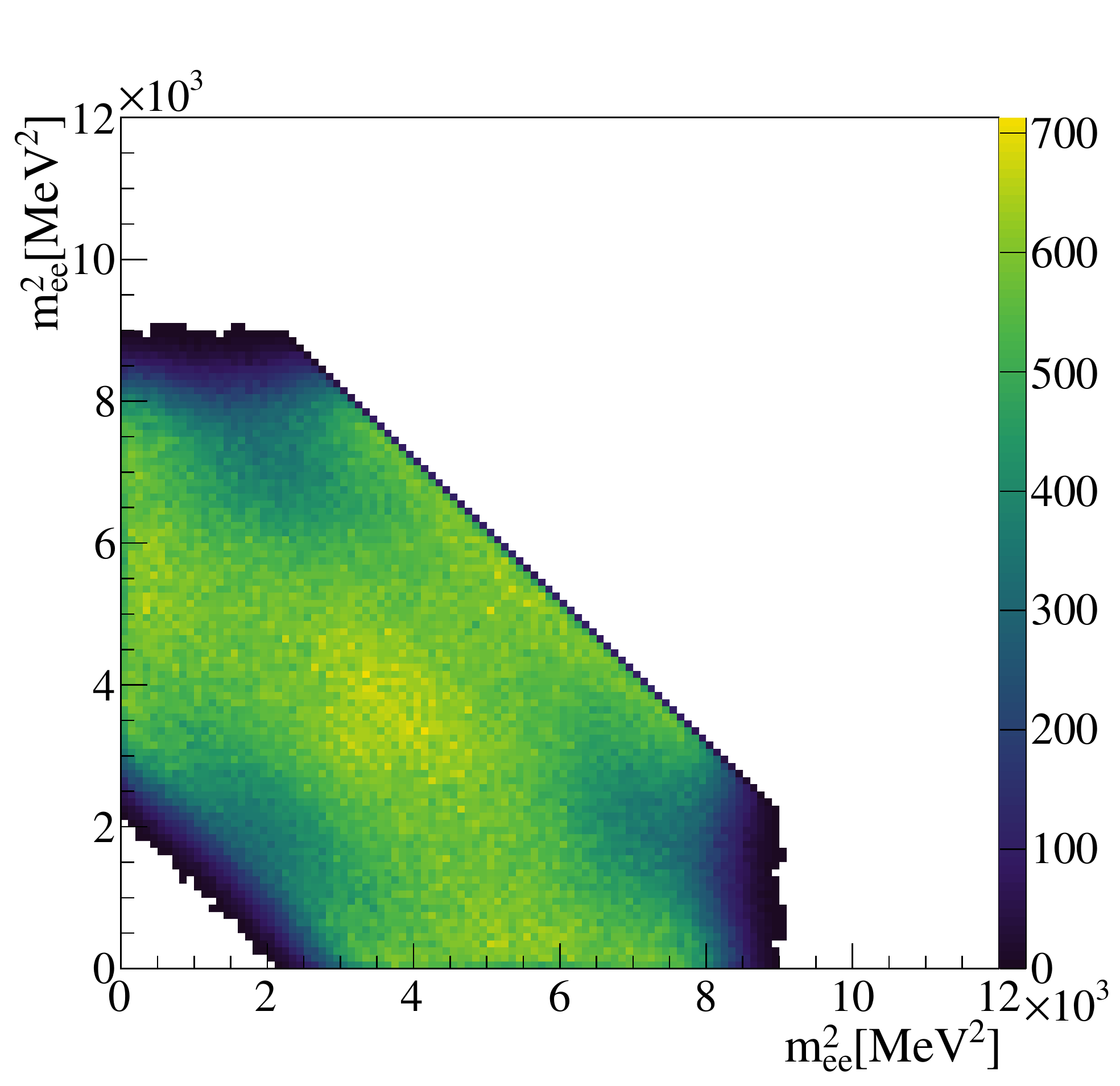}} \\
        \subfloat[Dipole operator. Note the logarithmic scale. ]{\includegraphics[width=0.3\textwidth]{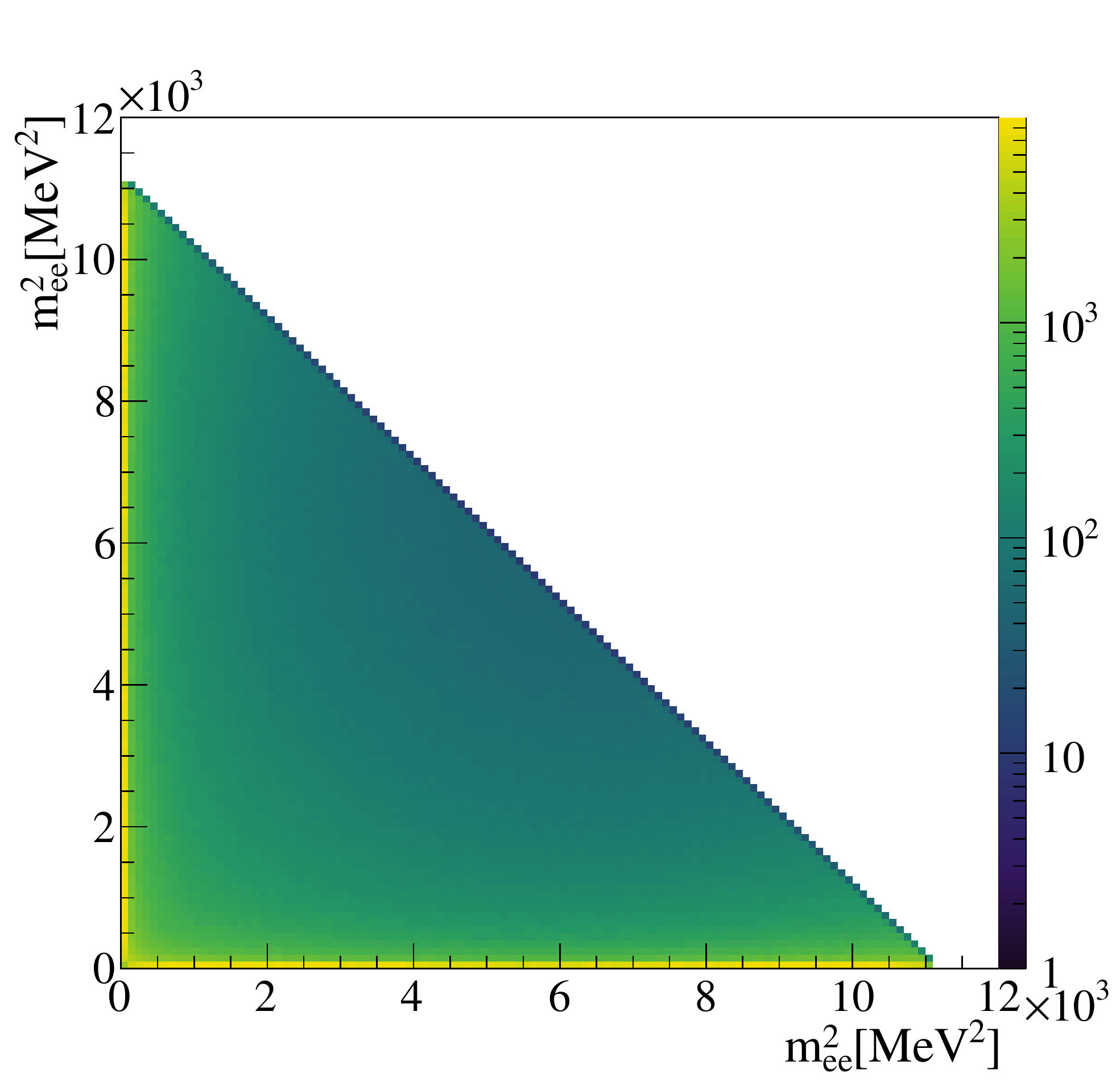}} \quad
        \subfloat[Scalar 4-fermion operator. ]{\includegraphics[width=0.3\textwidth]{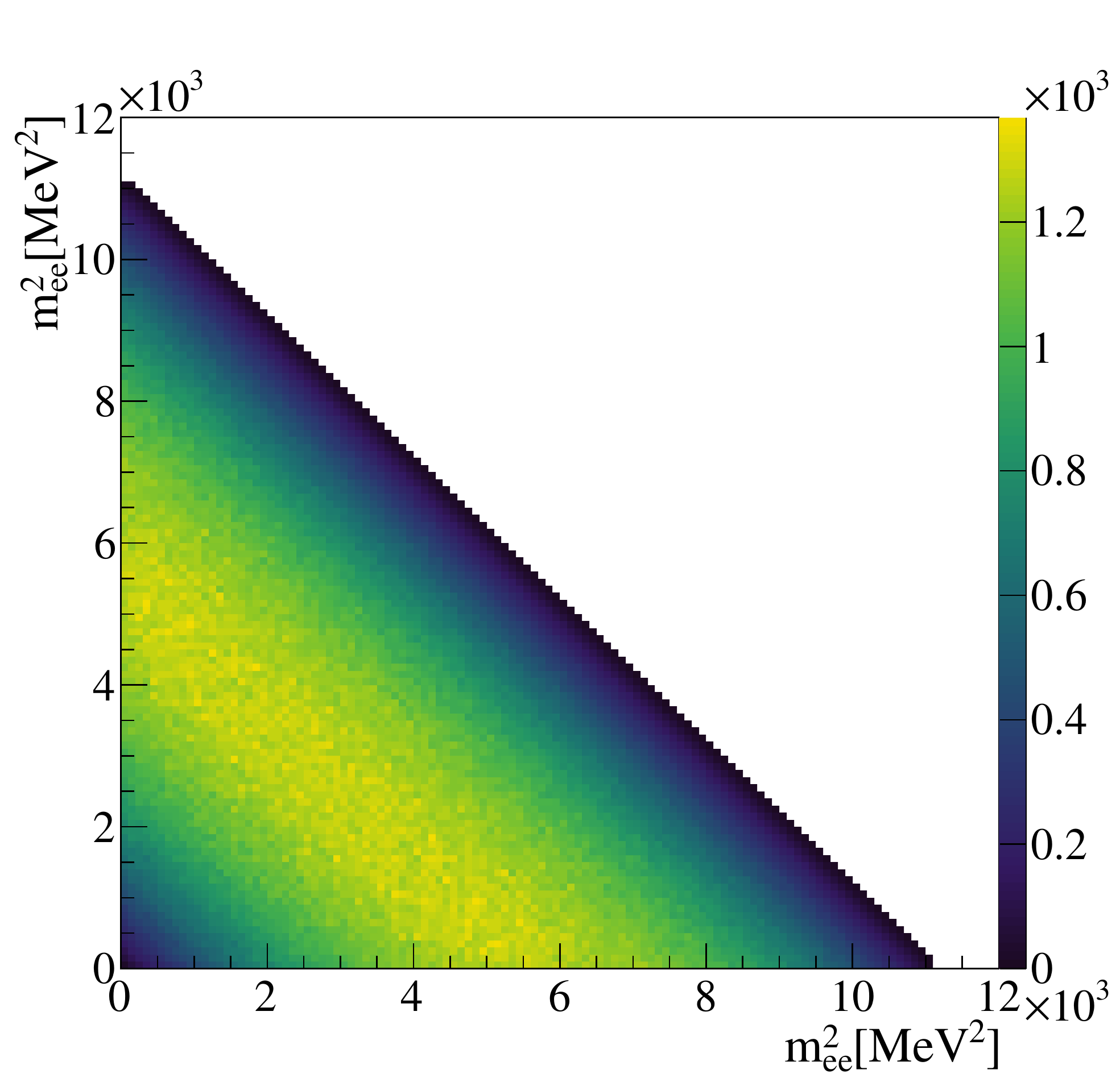}} \quad
        \subfloat[Vector 4-fermion operator. ]{\includegraphics[width=0.3\textwidth]{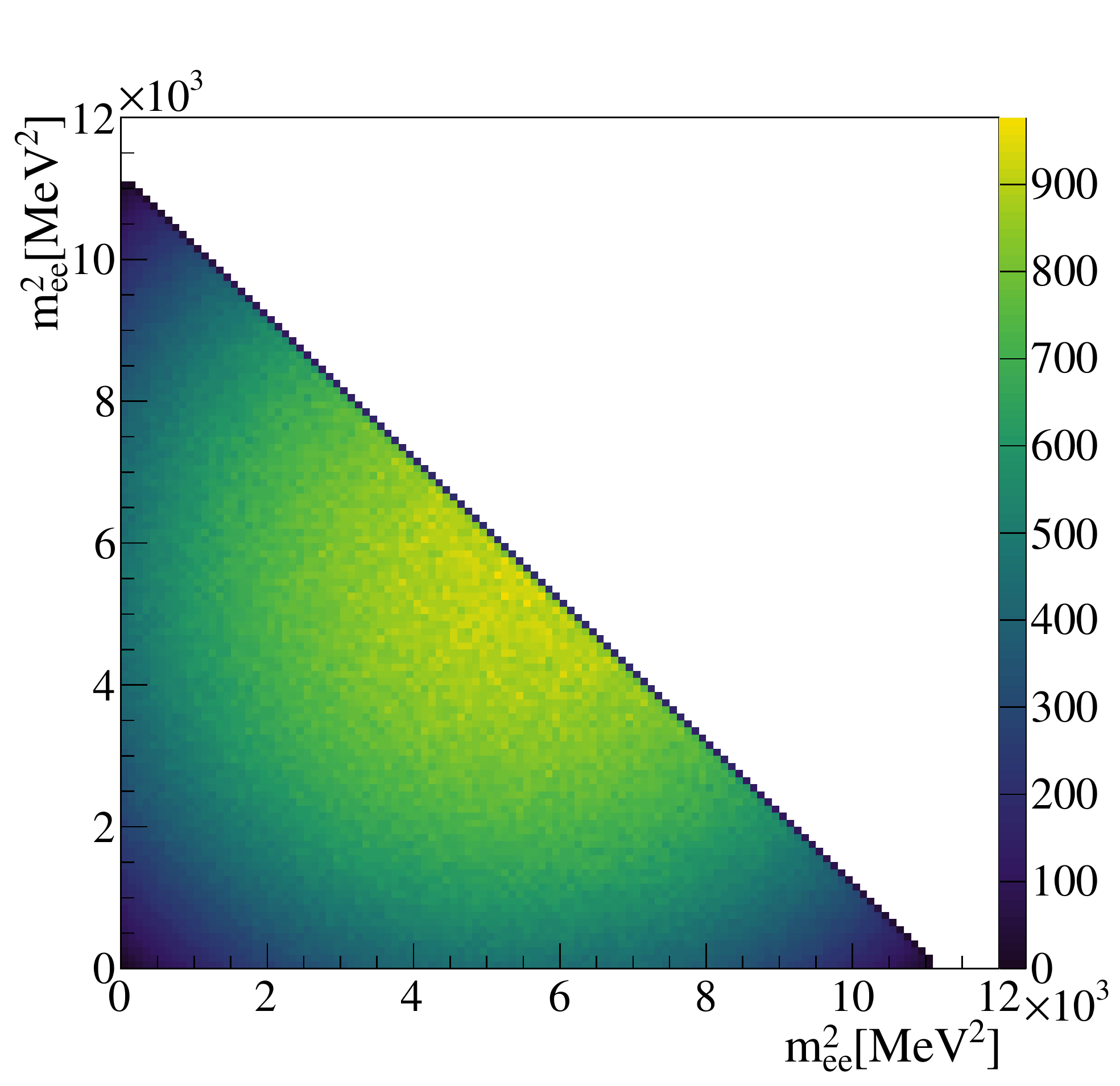}}
\caption{Dalitz plots of $e^+e^-$ pairs in simulated \mueee\ decays. In the effective
        theory approach, the operators are evaluated individually. }
\label{fig:Mu3eEFT}
\end{centering}
\end{figure}

In a baseline approach, the \mueee\ signal decay is implemented as a
three-body decay without any assumptions on the underlying physics. 
In the event selection, only vertices with three reconstructed
long tracks are kept because the momentum resolution of short tracks does not
suffice to suppress background from \mueeenunu. 
Constraints are applied on the quality of the vertex fit, the distance of the
reconstructed vertex to the target, the relative timing between the electron
candidates, as well as on the total momentum and invariant mass of the \mueee\
candidates. 
Figure~\ref{fig:Mu3e}\ shows the expected distribution of the invariant three
electron mass for background from \mueeenunu\ as well as for the \mueee\
signal at various branching fractions. 
With an expected number of 0.44 background events, background-free
operation is confirmed. 
The \phase\ experiment is estimated to be sensitive to branching fractions as
small as \num{5.2e-15}\ at \SI{90}{\percent}\ CL. \\
This approach however neglects the fact that the type of interaction which
mediates \mueee\ affects the kinematics of the decay and consequently the
efficiency to reconstruct signal events. 
The signal decay is therefore modelled with an effective field theory 
approach of dimension six. 
The operator basis and differential branching fraction are taken
from~\cite{Kuno:1999jp, Crivellin:2017rmk}. \\
In figure~\ref{fig:Mu3eEFT}, Dalitz plots of $e^+e^-$ pairs in \mueee\ are
shown for three-body decays and for various effective operators. 
The generated three-body decays are evenly distributed over the accessible
phase space as expected. The Dalitz plot of the reconstructed three-body
decays illustrates the acceptance of the detector. 
The corners are cut as a result of the minimum detectable $p_\text{T}$. 
Further structure stems from requiring long tracks only. \\
In the case of dipole operators, the virtual photon manifests in very small
invariant $e^+e^-$ masses which increases the probability to have at least one
decay particle with $p_\text{T}$ below the detection threshold. Thus, the
sensitivity is reduced compared to three-body decays with $\BR\geq\num{8.5e-15}$
at \SI{90}{\percent}\ CL. 
In the case 4-fermion operators on the contrary, the decay energy tends to be
more equally distributed amongst the decay particles which leads to an
increased detection efficiency. 
For all 4-fermion operators, comparable sensitivities of
$\BR\geq\num{4.6e-15}$ at \SI{90}{\percent}\ CL are estimated despite the
distinct decay distributions. \\
This effective operator approach further provides a base for the 
interpretation of observations or non-observations in the various lepton
flavour violation searches. 
A unique feature of \mueee\ searches is that in case of a discovery
conclusions on the type of interaction can already been drawn from decay
distributions such as Dalitz plots and angular distributions.

\subsection[Resonances in \ee]{Resonances in $\boldsymbol{\eemath}$}
\label{sec:darkphoton}
\begin{figure}
\begin{centering}
        \subfloat[Expected branching fraction limits at \mbox{\SI{90}{\percent}\ CL}. ]{\includegraphics[height=0.205\textheight]{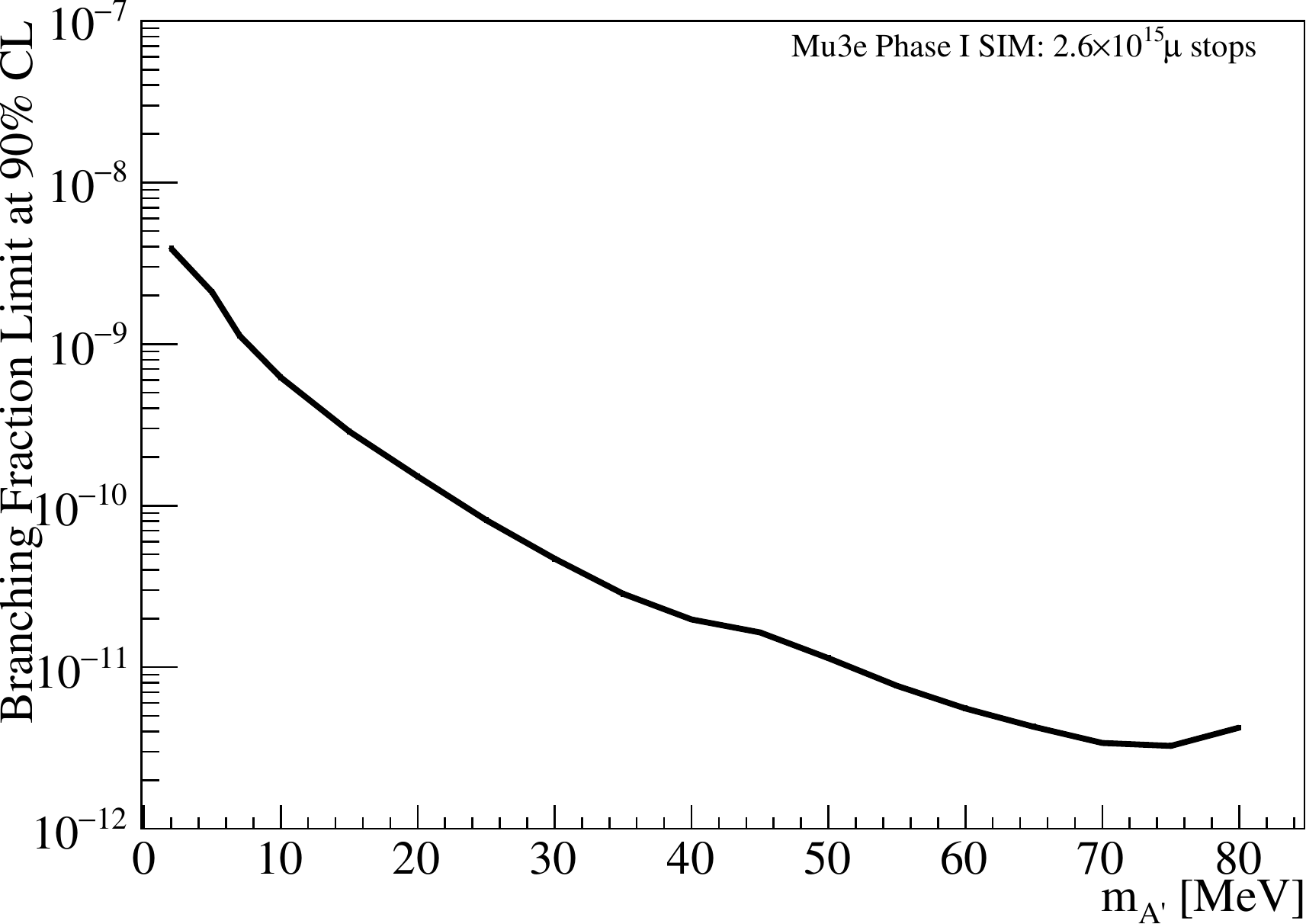}\label{fig:DPLimitsBR}} \quad
        \subfloat[Expected limits on the kinetic mixing parameter $\epsilon$ 
        at \mbox{\SI{90}{\percent}\ CL}. Adapted from~\cite{Ablikim:2017aab}. ]{\includegraphics[height=0.205\textheight]{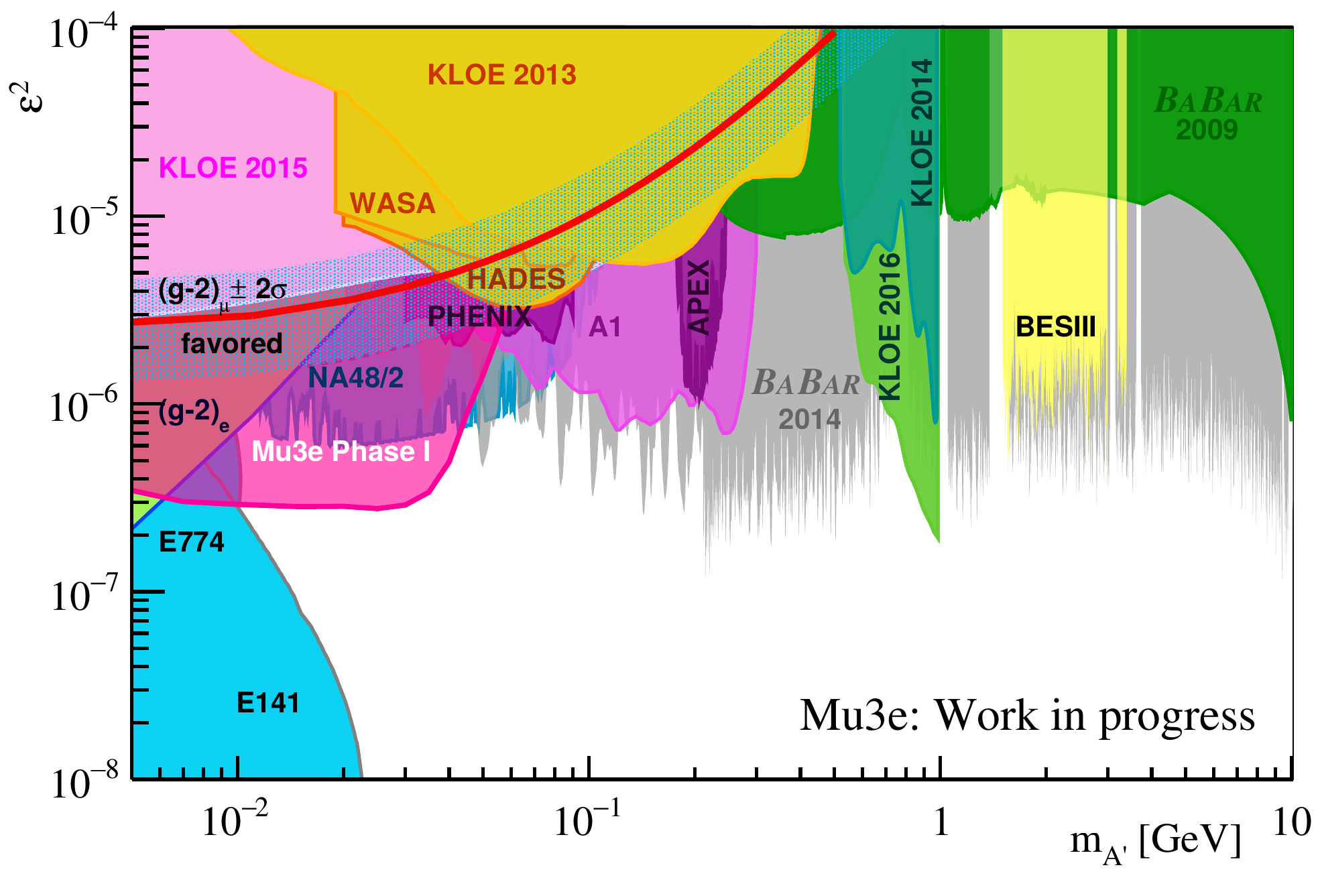}\label{fig:DPLimitsPara}} 
\caption{Sensitivity to prompt dark photon decays in \mueeenunu\ in
        the \phase\ Mu3e experiment. }
\label{fig:DPLimits}
\end{centering}
\end{figure}

In the search for \mueee\ events, also a unprecedented number of \mueeenunu\
decays will be collected. 
As no constraint on the invariant three electron mass is applied in online
reconstruction and event filtering, all \mueeenunu\ decays in the acceptance are
recorded. 
This data can be used for resonance searches in \ee\ pairs. \\
These resonances can be the signature of a dark photon -- a potential messenger to
a dark sector. 
The dark photon interacts via kinetic mixing with the \SM\ photon and $Z$ and
thus couples to the electro-magnetic current. 
If sufficiently light, it can be radiated in muon decays and decay promptly to
an \ee\ pair given the life time is short. \\
The background to \ee\ resonance searches is the same as for \mueee, but this
search cannot be performed background-free as the final state is the same as
in the \SM\ \mueeenunu\ decay. \\
Prompt dark photon decays in muon decays are generated with MadGraph5\_aMC@NLO
2.4.3~\cite{Alwall:2014hca}\ using a kinetic mixing
model~\cite{Echenard:2014lma}\ and fed into the detector simulation. 
Cuts on the quality of the vertex fit and distance of the reconstructed
vertex to the target are applied to reduce accidental background. 
Reconstructed short tracks are considered whenever the track could not be
extended to a long track. \\
\begin{figure}[t]
\begin{centering}
        \subfloat[Simulated background events. Both \ee\ combinations are
        considered. ]{\includegraphics[width=0.41\textwidth]{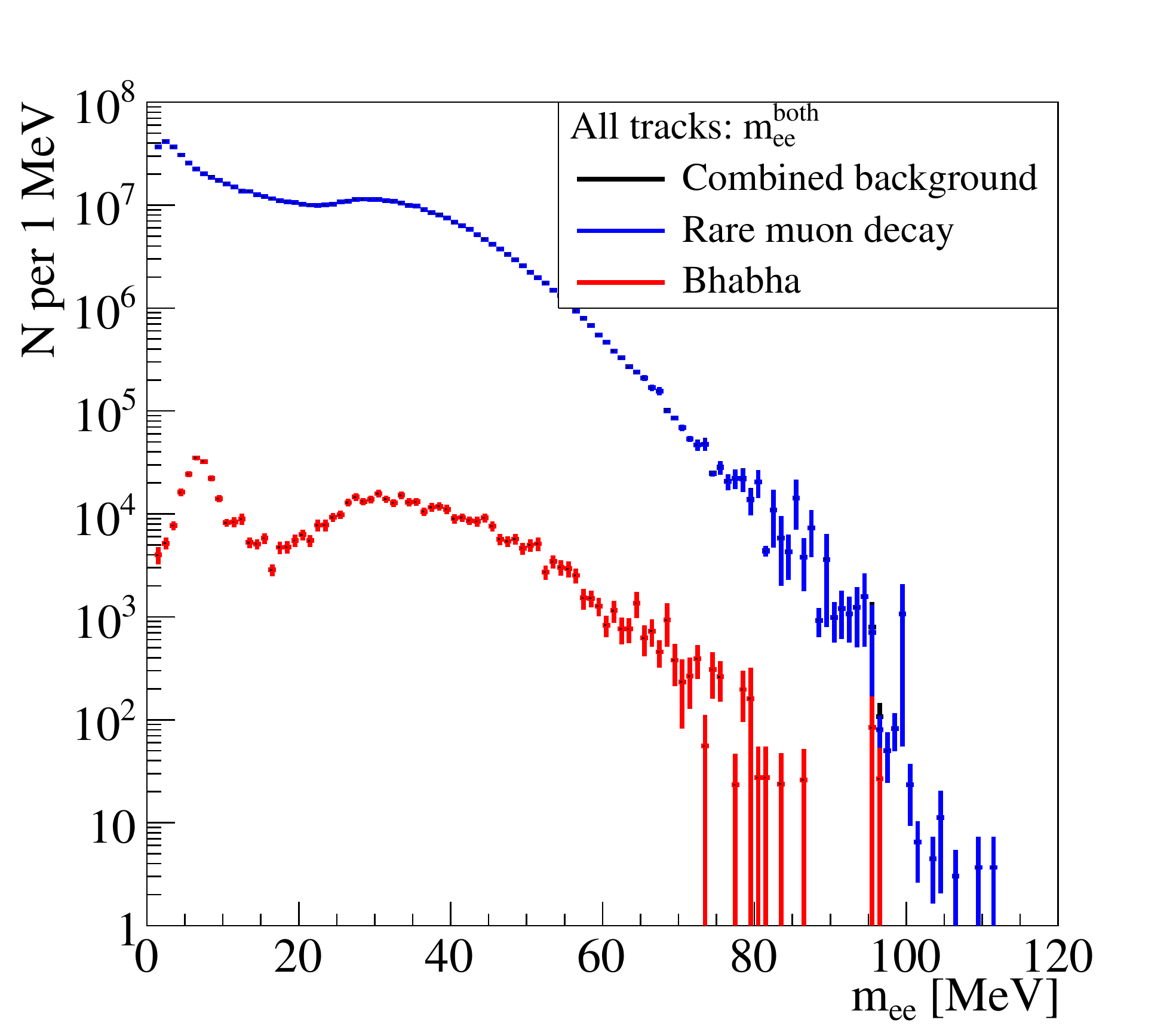}} \quad
        \subfloat[Simulated dark photon signal events. Both \ee\
        combinations are considered. ]{\includegraphics[width=0.41\textwidth]{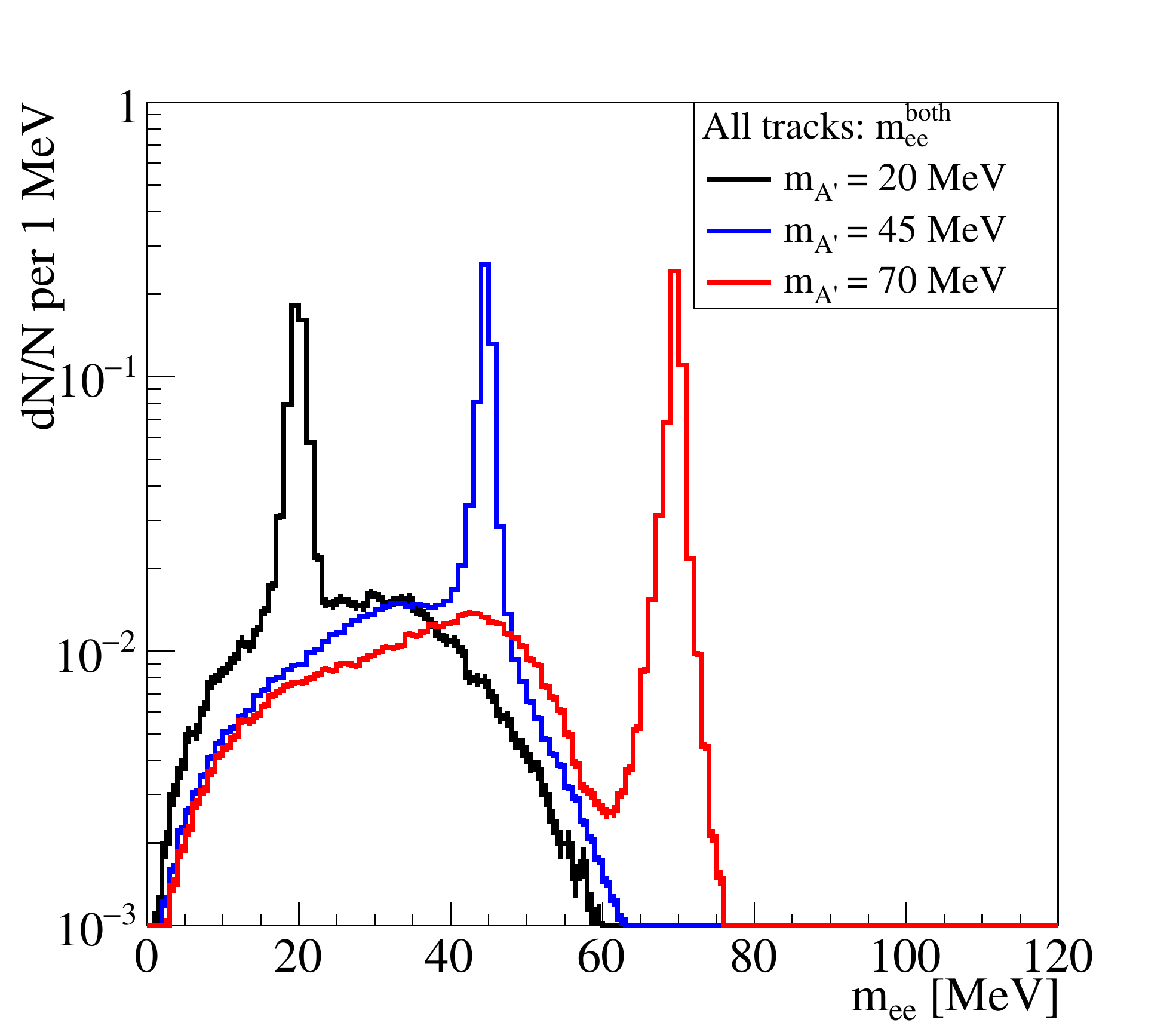}\label{fig:DPSigboth}} \\
        \subfloat[Simulated dark photon signal events. The \ee\
        combination with the lower invariant mass is shown. ]{\includegraphics[width=0.41\textwidth]{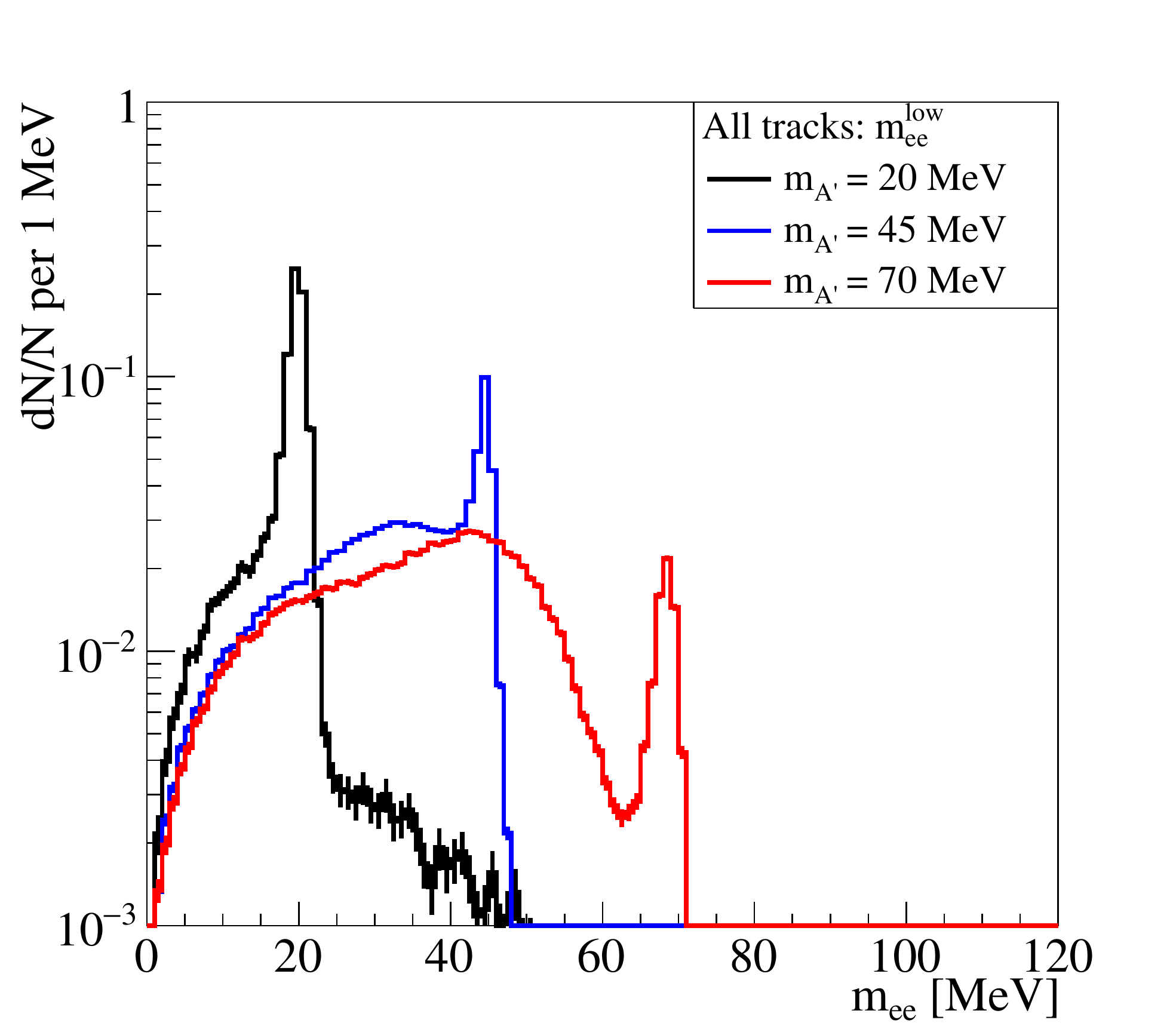}\label{fig:DPSiglow}} \quad
        \subfloat[Simulated dark photon signal events. The \ee\
        combination with the higher invariant mass is shown. ]{\includegraphics[width=0.41\textwidth]{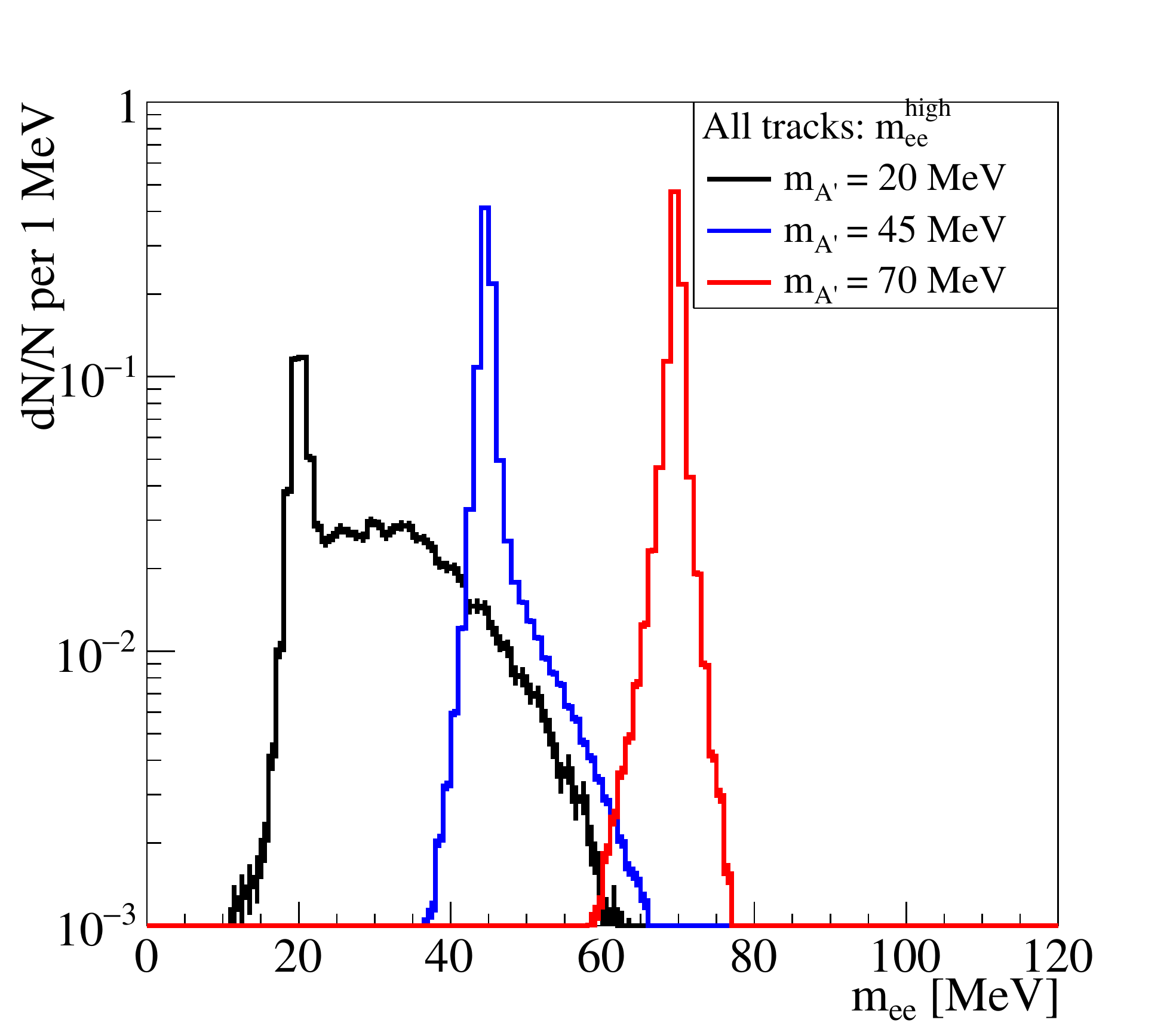}\label{fig:DPSighigh}} 
\caption{Distribution of the reconstructed invariant mass of \ee\ pairs
        for simulated \SM\ background and signal from prompt dark photon decays. }
\label{fig:DPSigBG}
\end{centering}
\end{figure}
The resulting distributions of the invariant \ee\ mass for signal and
background are shown in figure~\ref{fig:DPSigBG}. 
The background spectrum is smoothly declining in most of the mass range and is
dominated by \mueeenunu\ decays. Accidental background from combinations of
Bhabha scattering events and Michel decays contributes on average a factor of
800 less. Other types of accidental background are even less frequent and thus
negligible. \\
The signal distribution shows a clear peak around the dark photon mass in
addition to a broad distribution that stems from the second \ee\ combination
% which is not caused by the dark photon decay. 
of which the positron is from the muon decay and not from the dark photon decay. 
The sensitivity of the dark photon search is increased by choosing the lower
$m_\text{ee}$ pair at small dark photon masses and the higher $m_\text{ee}$
pair at high dark photon masses. The transition between the two regimes is at 
around \SI{45}{\MeV}\ dark photon mass. \newpage
\noindent The sensitivity is estimated in toy Monte Carlo studies. 
Dark photon masses $m_{\DPmath}$ up to \SI{80}{\MeV}\ are studied. 
For higher masses, the dark photon parameter space covered by Mu3e is already
excluded by existing experimental limits. \\
The expected limits on the branching fraction are shown in
figure~\ref{fig:DPLimitsBR}. These range from \num{5e-9}\
at \mbox{\SI{90}{\percent}\ CL}\ for small dark photon masses to \num{3e-12}\
at \mbox{\SI{90}{\percent}\ CL}\ for high $m_{\DPmath}$.
The branching fraction limits can be translated to limits on the kinetic
mixing parameter $\epsilon$ assuming the dark photon decays exclusively 
to \ee~\cite{Echenard:2014lma}. 
The results are shown in figure~\ref{fig:DPLimitsPara}. 
Due to the immense number of muon decays, the Mu3e experiment is 
capable to investigate a currently not covered parameter space. 

\subsection{Lepton Flavour Violating Two-Body Decays}
\label{sec:familon}
\begin{figure}
\begin{centering}
        \includegraphics[width=0.75\textwidth]{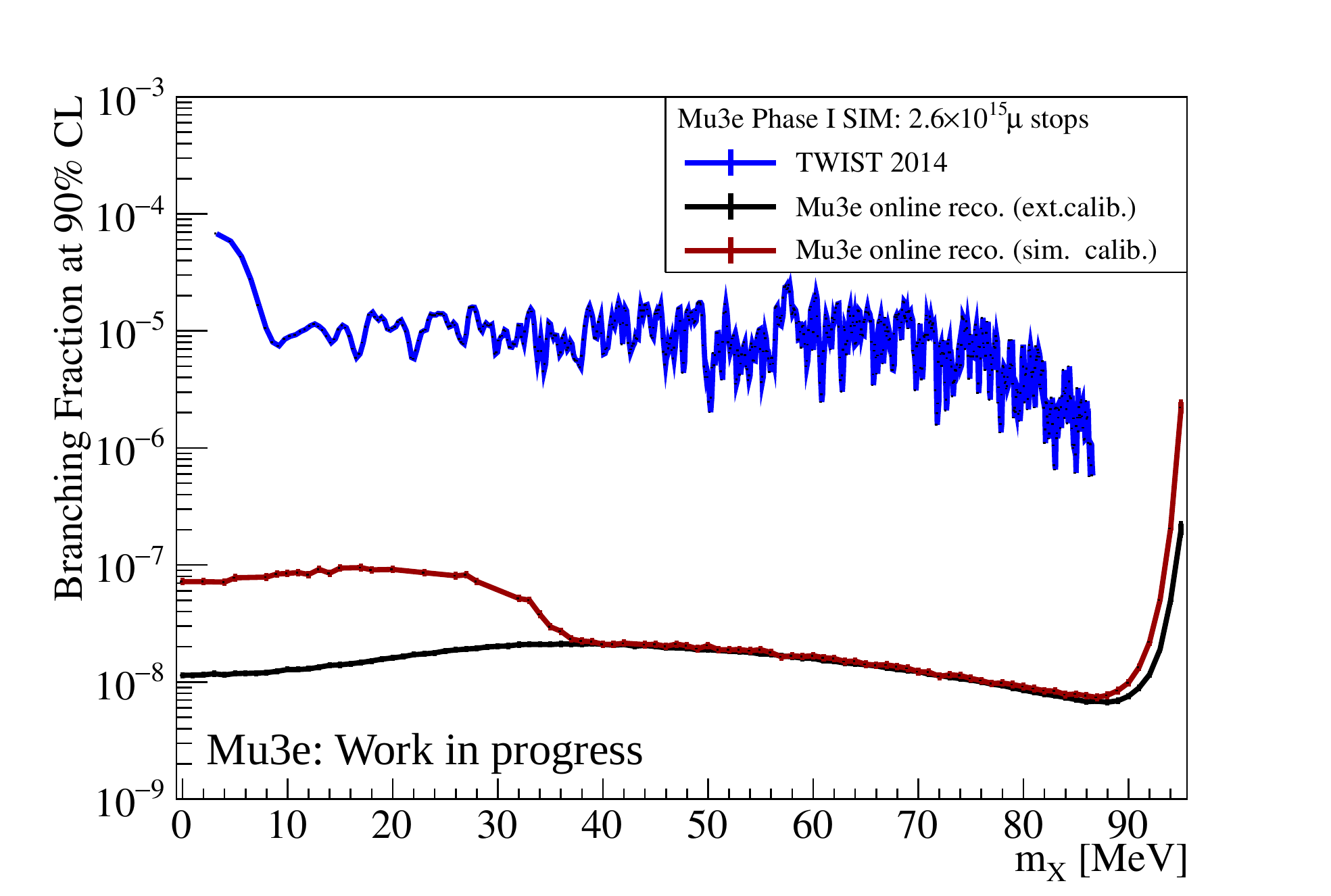}
\caption{Expected limits at \SI{90}{\percent}\ CL on the branching fraction of
\mueX\ in the \mbox{phase I}\ Mu3e experiment. 
The momentum calibration is either obtained from the same momentum spectrum
with a left out signal window (red line) or is assumed to be obtained from
another process such as Bhabha or Mott scattering (black line). 
Observed limits by the TWIST experiment are shown in blue~\cite{Bayes:2014lxz}.}
\label{fig:mueXBr}
\end{centering}
\end{figure}

A further channel that can be investigated with the Mu3e experiment is the
lepton flavour violating decay \mueX\ in which \familon\ denotes a neutral
light particle that is not detected in the experiment. 
Such a decay is motivated by the familon, a potential pseudo-Goldstone boson
arising from an additional broken flavour symmetry~\cite{Wilczek:1982rv}. \\
The signature of a \mueX\ decay is a monoenergetic positron whose energy is
determined by the mass of \familon. It can thus be identified by a narrow peak
on the smooth momentum spectrum of positrons from Standard Model muon
decays. \\ 
\begin{figure}
\begin{centering}
        \subfloat[Simulated background events. ]{\includegraphics[width=0.45\textwidth]{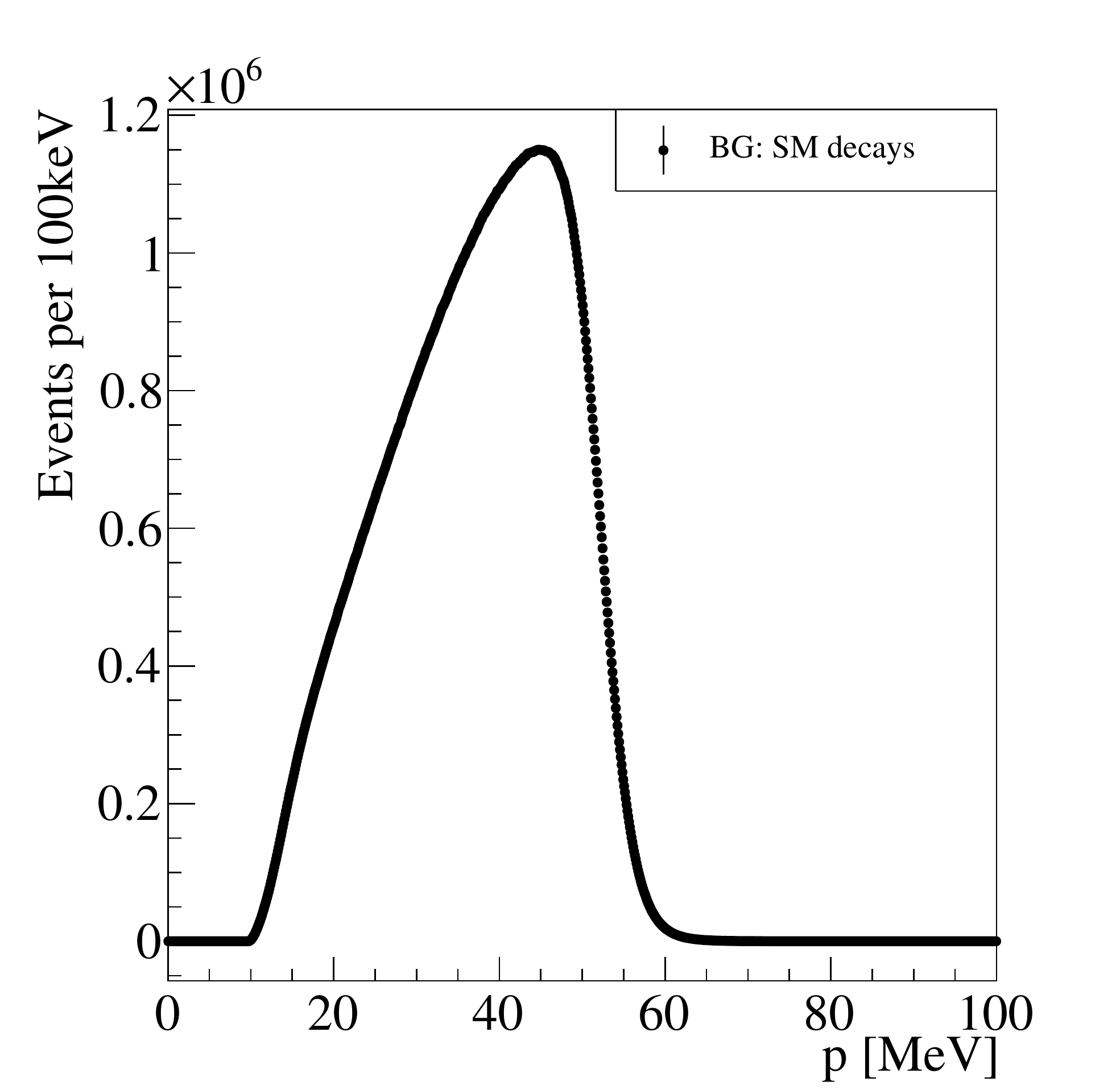}} \quad
        \subfloat[Simulated \mueX\ signal events. ]{\includegraphics[width=0.45\textwidth]{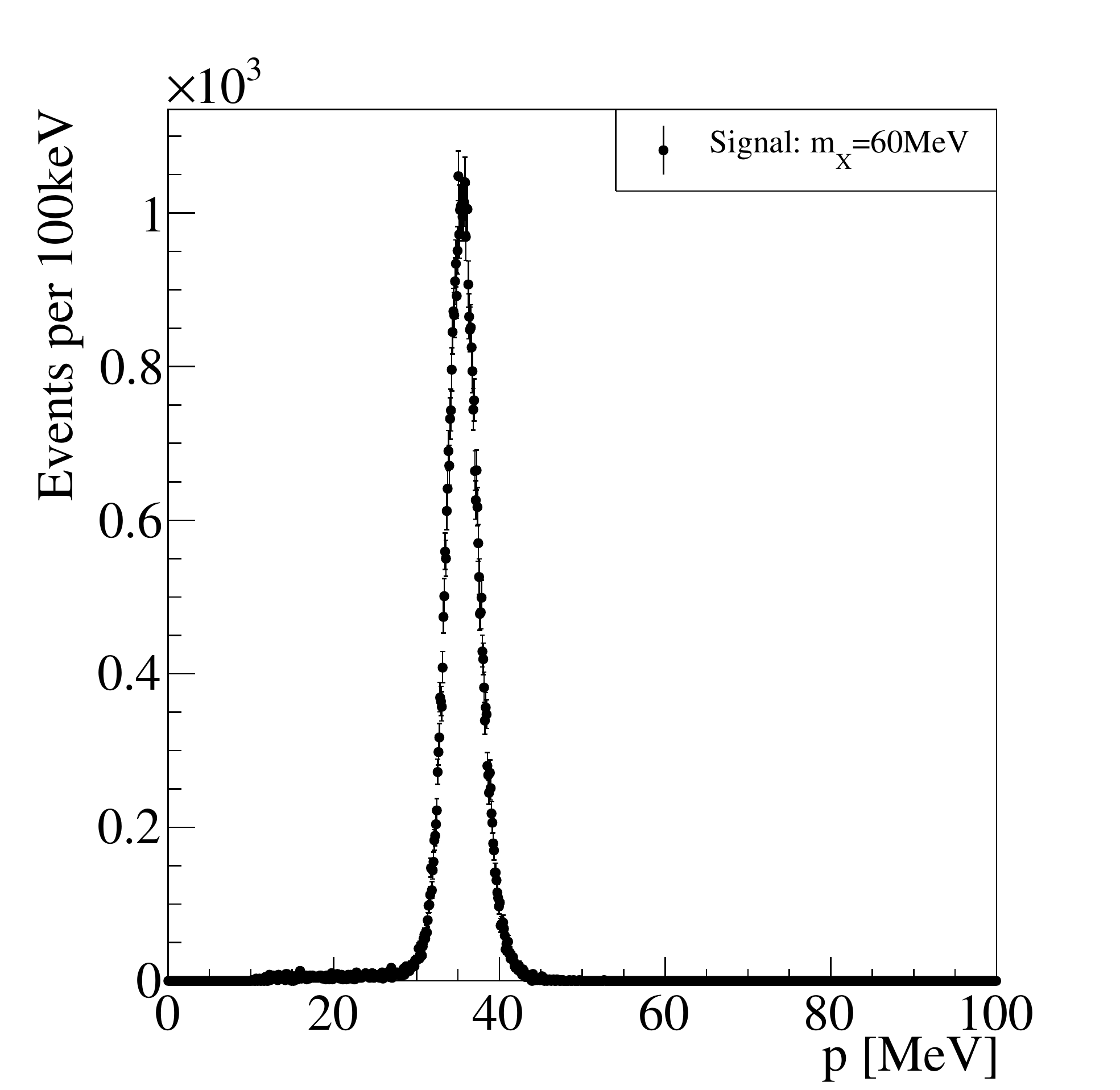}} 
\caption{Spectra of the reconstructed momentum of positrons from simulated
        \SM\ background and \mueX\ signal events. }
\label{fig:familonSigBG}
\end{centering}
\end{figure}
As the filter farm only selects events with \mueee\ candidates for data
storage, \mueX\ searches are performed on momentum histograms derived from
online reconstruction with the drawback of the non-optimal momentum
resolution of short tracks.  
The histograms are filled with tracks that are estimated to stem from the
target and pass selections on the quality of the track fit as well as on the
inclination angle. 
The latter selection efficiently removes tracks that stem from particles which
perform multiple loops in the detector. 
The resulting momentum spectra for signal and background are shown in
figure~\ref{fig:familonSigBG}. \\
The mass reach for \mueX\ searches is constrained by the minimum transverse
momentum of detectable positrons which allows to observe \familon\ with masses
of up to \SI{95}{\MeV}. 
Furthermore, the characteristic edge in the momentum spectrum of positrons
from the dominant Michel decay is the preferred means of absolute momentum
calibration in the baseline \mueee\ search. In this case, \mueX\ searches
cannot be conducted close to the Michel edge which affects massless and
light \familon . Alternative calibration methods relying on Bhabha or Mott
scattering are currently being studied. \\
The sensitivity in the first phase of the Mu3e experiment is estimated in toy
Monte Carlo studies using generated \mueX\ and Standard Model muon decays
reconstructed as short tracks.
For the results shown in figure~\ref{fig:mueXBr}, the momentum calibration is
either assumed to stem from an alternative approach such as Bhabha scattering,
or a window around the expected \mueX\ signal is left out while the calibration
is performed on the Michel spectrum. 
In the latter case, the sensitivity deviates from the first approach
as soon as the signal window reaches the Michel edge because the calibration
becomes less precise. \\
The latest limits on the \mueX\ branching fraction for massive \familon\ have
been derived by the TWIST experiment~\cite{Bayes:2014lxz}. Averaged over the
mass range, branching fractions larger \num{9e-6}\ are excluded
at \SI{90}{\percent}\ CL. 
Driven by the large number of muon decays, an improvement by a factor of 600
is estimated for the Mu3e experiment in \phase\ with expected branching
fraction limits in the order of \num{e-8}. 

\section{Conclusion}
\label{sec:conclusions}
The upcoming Mu3e experiment at PSI is going to search for the lepton flavour
violating decay \muposeee\ with an unprecedented sensitivity. 
Already in \phase, a branching fraction of \num{5.2e-15}\ can be measured or
excluded at \SI{90}{\percent}\ CL. 
Furthermore, conclusions on the type of underlying physics can be drawn from
the decay distributions in case of discovery. \\
In addition to \mueee\ searches, the experiment is also suited to investigate
other signatures of physics beyond the \SM. 
Substantial improvements can be expected in searches for decays of the type
\mueX\ for which branching fractions in the order of \num{e-8}\ can be
tested. 
Also in the case of dark photons a currently uncovered parameter space can
be studied.

\section*{Acknowledgements}
The presented work was supported by the Research Training Group `Particle
Physics Beyond the Standard Model' (GRK 1940) of the Deutsche
Forschungsgemeinschaft.

\bibliography{tau.bib}

%\nolinenumbers

\end{document}